\documentstyle[11pt]{article}
\newcommand{\be}{\begin{equation}}
\newcommand{\ee}{\end{equation}}
\newcommand{\bea}{\begin{eqnarray}}
\newcommand{\eea}{\end{eqnarray}}
\begin{document}
\centerline{\bf Domain Structures and Zig-Zag Patterns}
\centerline{\bf Modeled by a Fourth-Order Ginzburg-Landau Equation }
\vspace{1cm}
\centerline{David Raitt}
\centerline{Hermann Riecke}
\centerline{Department of Engineering Sciences and Applied Mathematics}
\centerline{Northwestern University}
\centerline{Evanston, IL 60208, USA}

\section{Introduction}

The formation of steady spatial structures in systems far from equilibrium has
been studied in great detail over the past years, the classical examples being
Rayleigh-B\'enard convection and Tayor vortex flow \cite{ch93}. In
quasi-one-dimensional geometries they usually share a common feature: the
stable structures that arise after the decay of transients are strictly
periodic in space and if they are weakly perturbed they relax diffusively
back to the periodic state. This relaxation has been investigated
experimentally
in various systems \cite{wa91} and
the results agree with theoretical results based on the phase diffusion
equation
\cite{rp87,lr90,pr91}. It describes the slow dynamics of the local phase, the
gradient of which is the local wave number. The structures are generally
stable over a range of wave numbers which is limited (at least) by the Eckhaus
instability which is characterized by
a vanishing of the diffusion coefficient. It leads through the destruction
(or creation)  of one or more unit cells, i.e. a roll pair in convection or
a vortex pair in Taylor vortex flow, to a new stable periodic state with a
different wave number.

Recently it has been pointed out that a vanishing of the diffusion coefficient
does not necessarily invoke the Eckhaus instability \cite{bd89,r90}. Instead,
under certain conditions, the system
can go to a state consisting of distinct domains in which the wave number has
different
values and which does {\it not} relax to a strictly periodic
structure. The domains are separated by domain walls in which
 the wave number changes rapidly.
This situation has been treated successfully using a higher-order
phase equation \cite{bd89,r90}.

Experimentally, inhomogeneous structures have been found in a variety of
systems. Most of them involve at least one time-dependent structure, e.g.
counterpropagating traveling waves (or spirals) \cite{mhaw90},
(turbulent) twist vortices amidst regular Taylor vortices \cite{ba86},
localized
traveling-wave pulses in binary-mixture convection \cite{bks90} and steady
Turing patterns amidst chemical traveling waves \cite{pdkwdb92}. Recently,
however, domain structures involving only steady convection rolls of two
different sizes have been observed in Rayleigh-B\'enard convection in a very
narrow channel \cite{hvdb92}. It has been speculated that these
structures may be related to the phase-diffusion mechanism discussed above
\cite{bd89}.

Within the framework of the phase equation domain structures consisting of
an array of domain walls are not stable due
to the attractive interaction between the walls. This leads to a coarsening of
the structures during which
domain walls annihilate each other. This dynamics is closely related to that
observed in
 spinodal decomposition of binary mixtures after quenches into the miscibility
gap
\cite{k93}. In general, the coarsening will eventually lead to periodic
structures with a constant wave number. Only if the boundary conditions
conserve the
total phase, i.e. require
 the total number of convection rolls, say,
to be constant, will the final state consist of a (single) pair of domain
walls.

Here we investigate the possibility of stable domain structures in the absence
of
phase conservation. To do so we study systems in which two periodic patterns
differing
only in their
wave numbers are equally likely to arise. In certain cases the
 competition between the two wave numbers can be described by
an extended Ginzburg-Landau equation. We study this equation with spatially
ramped
control parameter in order to allow the total phase to change and find that
even
in this general case domain
structures of varying sizes can be stable. This Ginzburg-Landau equation can
also be viewed
as a one-dimensional version of the Ginzburg-Landau equation for
two-dimensional patterns
in isotropic and anisotropic systems. The domain structures correspond then to
`zig-zag' patterns. In two dimensions phase conservation is less
common than in one dimension due to the possibility of focus singularities in
the pattern
which often arise at the boundaries \cite{cn84}. Our results may therefore also
shed some
additional light
on the behavior of the `zig-zag' patterns studied previously
\cite{pk86,bkkpwz90}.

\section{Numerical Simulation of Fourth-Order Ginzburg-Landau Equation}

\label{s:domain}
A situation in which the competition between two wave numbers can be
investigated
with relative ease is obtained if the neutral curve which marks the stability
boundary
of the basic state has two almost equal minima at (slightly) different wave
numbers.
For values of the control parameter $\Sigma$ slightly above these minima
the pattern can be described by a Ginzburg-Landau equation for the amplitude
$A$ \cite{p91},
\be
\partial_T A = D_2 \partial_X^2 A +
iD_3\partial_X^3 A - \partial_X^4 A + \Sigma A - |A|^2A,\label{e:GL4}
\ee
which gives for instance the vertical fluid velocity in convection {\em via}
\be
v_z(x,z,t)=\epsilon e^{iq_cx} A(X,T) f(z) + h.o.t. + c.c..
\label{e:texp}
\ee
For simplicity we assume in the following that the neutral curve has reflection
symmetry with respect to $q_c$ and set $D_3=0$. In this case the neutral curve
has two minima
 if $D_2$ is negative. The control parameter $\Sigma$ is proportional
to the temperature difference across the fluid layer, say.

In order to allow the number of rolls to change freely without any pinning by
the boundaries
we apply a subcritical spatial ramping to $\Sigma$. Thus $A$ goes to zero
before the
boundary of the system and the wave number selected by the ramp
\cite{kbjbc82,rp87}
corresponds
to one of the minima of the neutral curve. For the numerical simulation a
Crank-Nicholson
scheme with
$dx\approx 0.034$ is used.  Such a small grid spacing is required to reduce the
pinning
by the grid below the small attractive force between domain walls.

For slowly varying wave numbers $Q(\tau,\xi)\equiv \partial_\xi \phi$,
 the Ginzburg-Landau equation (\ref{e:GL4}) can be
reduced to an equation for the phase of the amplitude $A=R e^{i\phi}$,
\be
\partial_\tau\phi = (D + E \partial_\xi \phi + F (\partial_\xi
\phi)^2)\partial_\xi^2 \phi
 -G \partial_\xi^4 \phi. \label{e:phase2}
\ee
Domain structures arise for negative values of $D$.
For large values of $\Sigma$ wave-number gradients across domain walls are
sufficiently
small to be described by (\ref{e:phase2}), and within this framework the
interaction between
domain walls is purely attractive.
Thus, in the presence of a subcritical ramp
domain structures are expected to evolve to a strictly periodic pattern. This
is confirmed
for $\Sigma=100$ and $D_2=-1$, using initial conditions
with three domains, i.e. a region of low wave number between two regions
of high wave number.

\begin{figure}[hbt]
\begin{picture}(420,250)(0,0)
\put(100,0) {\includegraphics{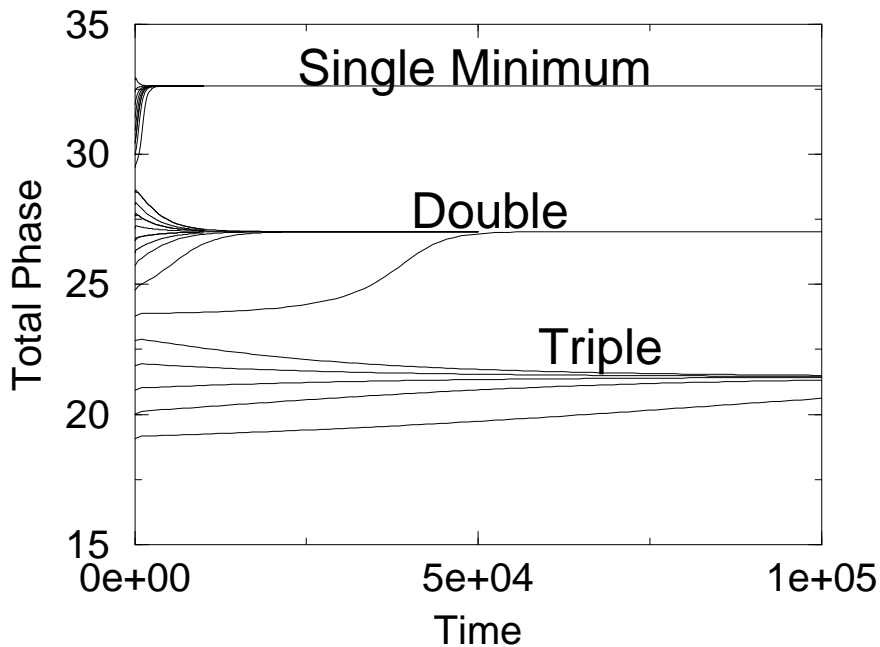}}
\end{picture}
\caption{Evolution of the total phase without phase conservation.  The
corresponding
final states are shown in fig.\protect{\ref{f:3states}}.
\protect{\label{f:S1phase}}}
\end{figure}

\begin{figure}[hbt]
\begin{picture}(420,330)(0,0)
\put(100,0) {\includegraphics{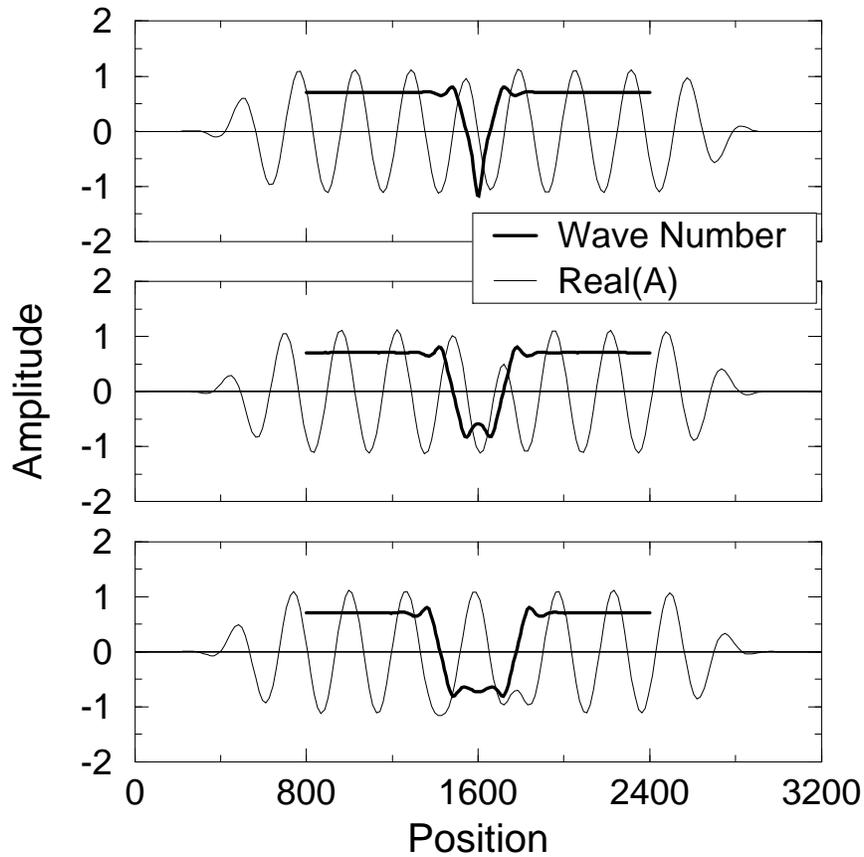}}
\end{picture}
\caption{Coexisting stable final states obtained from the time evolution shown
in
fig.\protect{\ref{f:S1phase}}.  The thick line gives the local wave number in
the bulk ($800<i<2400, dx=0.034$).  Note that the amplitude goes to zero toward
the
boundaries due to the subcritical ramping.
\protect{\label{f:3states}}}
\end{figure}

For $\Sigma=1$, on the other hand, the behavior is quite different, as
shown in fig.\ref{f:S1phase}.  It gives the total phase $\int Q\,dX$
in the system (between the
ramps) as a function of time for different initial conditions which are
obtained by
changing the width of the central (low wave number) domain. While for large
$\Sigma$
 the system evolves to the same periodic state independent of the initial total
phase,
it converges here to one of several final states with different total phase.
These states are pictured in fig.\ref{f:3states}. They differ in the width of
the
central domain and, most strikingly, their local wave number exhibits
oscillatory
behavior in space.

 The possibility for spatial oscillations in the wave number can be seen in a
linear stability
analysis of the periodic state. It shows that solutions which approach a
periodic
state for $X \rightarrow \pm \infty$ can do so in an oscillatory manner if the
fourth
derivative is present. The purely attractive interaction between
domain walls within the phase equation, and therefore also for large $\Sigma$,
is due to the monotonic behavior of the wave number across
the wall. The non-monotonic behavior found for smaller $\Sigma$ therefore
strongly suggests an oscillatory contribution to the interaction which would
explain the stability of bound pairs of domain walls.

\begin{figure}[hbt]
\begin{picture}(420,250)(0,0)
\put(100,0) {\includegraphics{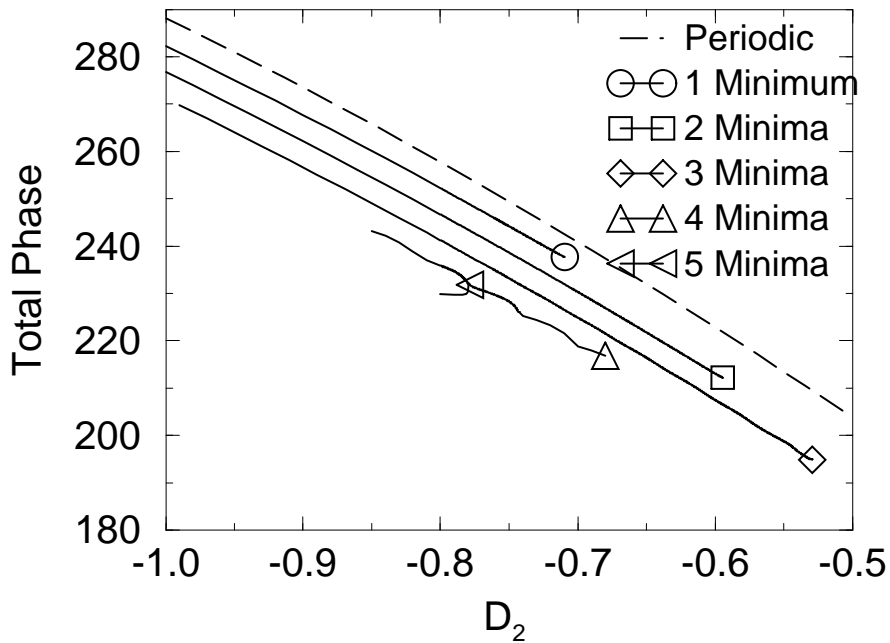}}
\end{picture}
\caption{Regime of existence of bound pairs of domain walls for $\Sigma =1$.
The calculation
was performed with 12800 points and a grid spacing of 0.034. The ramped part is
400 points
wide on each side.
\protect{\label{f:d2var}}}
\end{figure}

To study the transition between the two regimes one could investigate the
persistence of the
domain structures when increasing $\Sigma$. Since the scale for $\Sigma$ is set
by $D_2$,
which determines the depth of the wells in
the neutral curve, increasing $\Sigma$ is equivalent to making $D_2$ less
negative. In fact,
one of the 2 parameters could be scaled away ($D_2 \rightarrow D_2/|\Sigma|,
\Sigma
\rightarrow \pm 1$).
We therefore investigate the persistence of bound pairs by changing $D_2$
rather than
$\Sigma$ since this also sheds some light on the stability of the
two-dimensional
patterns discussed below. The result is shown in fig.\ref{f:d2var}. It gives
the total
phase within the unramped region of various states as a function of $D_2$. For
these simulations the system has been chosen much larger in order to avoid
interactions between the
domain walls and the ramps.
The periodic state is indicated by a dashed line. The different
 symbols denote where the bound pairs
with 1 to 5 minima in the local wave number, respectively, disappear.  While
the states with
1 to 3 minima jump to the periodic solution above that value of $D_2$, the
4-state jumps to the 3-state and the 5-state merges with the 4-state. The
5-state is
relatively hard to track due to
the rapid decay of the oscillations away from the domain walls.
For $D_2 < -0.8$ all six states
 coexist stably. In this regime arrays of domain walls should be possible in
which the widths of successive domains alternate chaotically \cite{cer87}.
Clearly, none of the bound pairs investigated persists all the way to $D_2=0$.
This corresponds to the previous result that for large $\Sigma$ the phase
equation becomes valid.
Strikingly,
the value of $D_2$ at which the solutions cease to exist is not a monotonic
function
of the number of minima in the wave number. Rather, the solution branch
corresponding to
the state with 3 minima is the last non-periodic state to disappear.
If one were to start with a chaotic array of domain walls and to increase
$D_2$, fig.(\ref{f:d2var}) suggests that domains of
various widths would successively be eliminated
beginning with the longest and the shortest ones until
only domains with 3 minima in the wave number remain.
Note that these simulations only address the interaction between two domain
walls. In a general array with many domains the regime of existence may be
therefore
somewhat different.

\section{Application to Two Dimensional Systems: Zig-Zags}

The above results are expected to capture also certain aspects of patterns
extended in
two dimensions. A particularly clear example is given by electro-hydrodynamic
convection
(EHC) in nematic liquid crystals. In the nematic phase the rod-like molecules
are
predominantly oriented in one direction rendering the fluid anisotropic.
Depending on
parameters, convection arises
in the form of rolls perpendicular or oblique to that preferred direction.
 Due to reflection symmetry, the oblique rolls can have wave number $(Q,P)$ or
$(Q,-P)$.
Often domains of oblique rolls with
opposite orientation are observed to coexist, leading to `zig-zag'
patterns. They are characterized by sharp transitions between the
domains with the two orientations.

Quite generally, in isotropic systems, as for instance in Rayleigh-B\'enard
convection (RBC),
straight roll patterns exhibit a secondary instability which leads to an
undulatory
deformation
of the rolls. It arises for small wave numbers and tends to increase the local
wave number.
In the course of the nonlinear evolution the undulations can grow sharper and
form
zig-zag-type patterns which may in fact be stable, as recently shown in
simulations of chemical Turing patterns \cite{db92}.

Near onset both systems can be described by suitable Ginzburg-Landau equations
which
differ in their scaling of the spatial variables. In the isotropic case one
obtains
\cite{nws69}
\be
\partial_T A = - (i\partial_X + \partial_Y^2)^2 A + \lambda A - |A|^2 A,
\label{e:iso}
\ee
whereas the anisotropic case yields \cite{pk86}
\be
\partial_T A = ( \partial_X^2  - i Z \partial_X \partial_Y^2  + W \partial_Y^2
- \partial_Y^4 ) A + \lambda A - |A|^2 A.
\label{e:aniso}
\ee
In the anisotropic case the normal/oblique transition occurs at $W=-ZQ$.

Focussing on solutions which are strictly periodic in $X$, $A = A_1(Y,T)
e^{iQX}$,
one obtains eq.(\ref{e:GL4}) for $A_1$,
\be
\partial_T A_1 = D_2 \partial_Y^2 A_1 - \partial_Y^4 A_1 + \Sigma A - |A_1|^2
A_1,
\label{e:gl1}
\ee
with
\bea
D_2=Q, \ \ \Sigma = \lambda - Q^2 \ \ \mbox{for (\ref{e:iso})\ }\\
D_2=W+ZQ, \ \ \Sigma = \lambda - Q^2 \ \ \mbox{for (\ref{e:aniso}).}
\eea
Note that $Y$ and $P$ in the two-dimensional systems correspond to $X$ and
$Q$ in eq.(\ref{e:GL4}). Thus, the domain structures discussed
in sec.\ref{s:domain}
are one-dimensional analogs of zig-zag patterns. The general
stability of zig-zag patterns in anisotropic systems has been studied by Pesch
and
Kramer \cite{pk86} and by Bodenschatz {\it et al.} \cite{bkkpwz90} who find
islands of
stability for the zig-zag states. They are presumably related to the discrete
set of
domains with different widths discussed above. Of course, the stability
of the zig-zags may be limited by two-dimensional instabilities which may
preempt some
 of the transitions discussed here. The interaction
of domain walls separating the two kinds of oblique rolls has, however,
not been discussed in detail.

The straight (normal) roll state corresponds to a solution with constant $A_1$.
Reducing its
wave number is equivalent to decreasing $D_2$, which goes through zero at the
onset of the
zig-zag instability. It leads to the growth of undulatory deformations of the
rolls
which correspond to periodic variations of the local wave number in
eq.(\ref{e:GL4}).
Based on the simulations of that equation one would expect that $D_2$ has to be
sufficiently negative for the resulting zig-zag structure to persist.
Therefore, close to the onset of the zig-zag instability the coarsening
dynamics predicted by the phase equation (\ref{e:phase2}) will prevail, and the
expected final state would be
a  pattern of straight (oblique) rolls of either sign if the phase is not
conserved.
 For deeper quenches into
 the unstable regime\footnote{Note $|D_2/\Sigma|\rightarrow \infty$ at the
neutral curve
$\Sigma = \lambda-Q^2$}, however, the undulations are expected to lock into
each other due to
the oscillatory interaction and form an array of domain walls, i.e. an array of
zigs and zags, which again could be spatially chaotic.

\section{Conclusion}

The competition between patterns differing only in their wave number can lead
to
complex spatial patterns in which the interaction between the walls separating
the domains with different wave number plays an important roll. Previously,
we investigated them for conditions
under which the total phase, i.e. the number of convection rolls in the system,
is conserved
\cite{rr93}. There we found a very rich bifurcation structure which originates
 from interactions with other modes arising from the Eckhaus instability.

In experiments the phase is not always conserved.
 In the thin layers used in electro-convection
of liquid crystals, for instance, inhomogeneities in the layer thickness are
often
strong enough that convection arises in large patches which are separated by
non-convecting
regions. In isotropic two-dimensional systems focus singularities are expected
to arise
in the corners of the container which also allow rolls to disappear smoothly
without
strong pinning \cite{cn84}. Here we therefore studied domain structures in the
presence
of spatial ramps. We found that they can exist even without phase conservation
and we
attribute
this to the oscillatory behavior of the local wave number which should lead to
an
oscillatory interaction between adjacent domain walls. The distance between the
walls
is discretized accordingly. Close to the onset of the zig-zag instability in
two-dimensional systems, these oscillations are small; to stabilize zig-zag
patterns
one therefore has to quench deeper into the unstable regime.

This work has been supported by grants from NSF/AFOSR (DMS-9020289) and DOE
(DE-FG02-92ER14303).

\end{document}